\documentstyle[twocolumn,aps]{revtex}
\begin{document}

\title{Higher dimensional Chern-Simons supergravity} 

\author{M\'aximo Ba\~nados$^1$, Ricardo Troncoso$^{1,2}$ and Jorge
Zanelli$^{1,3}$} 

\address{$^1$Centro de Estudios Cient\'{\i}ficos de Santiago, Casilla
16443, Santiago, Chile \\ 
$^2$Departamento de F\'{\i}sica, Facultad de Ciencias
F\'{\i}sicas y Matem\'aticas, Universidad de Chile, Casilla 487-3
Santiago, Chile \\
$^3$Departamento de F\'{\i}sica, Facultad de Ciencia,
Universidad de Santiago de Chile, Casilla 307, Santiago 2, Chile}

\maketitle

\begin{abstract}  
A Chern-Simons action for supergravity in
odd-dimensional spacetimes is proposed. For all odd dimensions, the
local symmetry group is a non trivial supersymmetric extension of
the Poincar\'e group. In $2+1$ dimensions the gauge group reduces to
super-Poincar\'e, while for $D=5$ it is super-Poincar\'e with a central
charge.  In general,  the extension is obtained by   the addition of a
1-form field which transforms as an antisymmetric fifth-rank  tensor
under Lorentz rotations. Since the Lagrangian is a Chern-Simons density
for the supergroup, the supersymmetry algebra closes off shell without
the need of auxiliary fields.  


\end{abstract}

\section{Introduction}

In the search for an unified theory of all interactions including
gravity, higher dimensional models have become standard in theoretical
physics. Two important examples are $D=10$
superstrings\cite{Green-Schwarz-Witten} and $D=11$
supergravity\cite{Cremmer-Julia-Scherk} which give rise, upon
dimensional reduction, to interesting effective models in four
dimensions.  In the study of higher --and lower-- dimensional models,
Chern-Simons densities play an important role. Although Chern-Simons
forms first appeared in the physics literature in the context of
anomalies, it is now clear that they have an intrinsic value as
dynamical theories in their own right.  For example, pure gravity and
extended supergravity in three dimensions are Chern-Simons theories for
the groups $SO(2,2)$ \cite{Witten88} and $OSp(2|p)\otimes OSp(2|q)$ 
\cite{Achucarro-Townsend}, respectively.  Similarly, in five dimensions,
supergravity can be written as a  Chern-Simons action for the
super-group $SU(2,2|N)$ \cite{Chamseddine2}. Chern-Simons forms also
provide a simple description for pure gravity in all odd-dimensional
spacetimes\cite{Chamseddine1}. The equations of motion of these theories
possess black hole solutions \cite{BTZ2} generalizing those found in 2+1
dimensions\cite{BTZ}. 

The Chern-Simons Lagrangian is constructed as follows. Let $G_A$ a basis
for the Lie algebra of a given (super)group {\bf G}. Let $A=A^A G_A$ be
the connection for {\bf G} and 
$F=dA + A\mbox{\tiny $\wedge$} A=F^A G_A$ its curvature
2-form. The Chern-Simons Lagrangian, $L$, is a ($2n+1$)-form whose
exterior derivative (defined in $2n+2$ dimensions) satisfies: 
\begin{eqnarray}
  dL &=& <F\mbox{\tiny $\wedge$} \cdots 
                \mbox{\tiny $\wedge$} F >, \nonumber\\
     &=& g_{A_1...A_{n+1}} F^{A_1} 
        \mbox{\tiny $\wedge$} ... \mbox{\tiny $\wedge$} F^{A_{n+1}}
\label{CSL} 
\end{eqnarray}  
where $g_{A_1...A_{n+1}}\equiv <G_{A_1},...,G_{A_{n+1}}>$ is completely
symmetric and satisfies the invariance condition 
\begin{equation}
\nabla g_{A_1...A_{n+1}}=0.
\label{Dg}
\end{equation}
[In the case of a supergroup, the invariant tensor
should have the corresponding (anti-) symmetry properties.]

For a given $n$, the above condition may have no solutions. Indeed, 
(\ref{Dg}) imposes strong restrictions on the group.  As we shall see
below, for $D>3$, one needs to enlarge the bosonic sector of the theory
in order to produce a supersymmetric extension of Chern-Simons gravity
that contains local Poincar\'e invariance.

It is direct to prove from (\ref{CSL}) that, up to a
total derivative, $L$ is invariant under the gauge transformation
\begin{equation} 
\delta_\lambda A = \nabla \lambda, 
\label{CSL/sym}  
\end{equation}   
where $\lambda$ is an arbitrary zero-form parameter and
$\nabla\lambda = d\lambda + [A,\lambda]$ . 
If $\delta_\lambda$ and $\delta_\eta$ are two
trasformations, with parameters $\lambda$ and $\eta$ respectively, then 
\begin{equation} 
[\delta_\lambda, \delta_\eta ] A =
\delta_{[\lambda,\eta]} A. 
\label{CSL/algebra}  
\end{equation}
   
The three dimensional supergravity studied by Ach\'ucarro and Townsend
\cite{Achucarro-Townsend}, as well as the five-dimensional theory
studied by Chamseddine \cite{Chamseddine2}, are Chern-Simons theories in
the sense described above. Their supersymmetry transformations can be
written in the form (\ref{CSL/sym}) and therefore the supersymmetry
algebra closes off-shell without the need of auxiliary fields. 

It goes without saying that, appart from the invariance under
(\ref{CSL/sym}), the Chern-Simons action is also invariant  under
diffeomorphisms. In 2+1 dimensions that symmetry is not independent from
the local gauge group because, as a consequence of the equations of
motion, the connection is locally flat.  This means that, given two
configurations that differ by a diffeomorphism, there always exist a 
gauge transformation that deforms one into the other.  In the
canonical formalism this is reflected by the absence of new {\em
independent} constraints associated with diffeomorphisms. 

In dimensions greater than three, however, the Chern-Simons equations of
motion do not impose the flatness condition and therefore the
diffeomorphism invariance {\em is} an independent symmetry giving rise
to independent constraints in the canonical formalism (see \cite{BGH}
for more details on this point). For our purposes here it is enough to
observe that the gauge trasformations (\ref{CSL/sym}) form a subgroup of
the whole symmetry group.  

It is the purpose of this paper to show that the above scheme can be
extended to supergravity in all odd dimensions, provided one chooses
the bosonic Lagrangian in an appropriate way.  It turns out that for
$D>3$, the Lagrangian for the bosonic sector is not Hilbert's. Rather,
the correct Lagrangian is non-linear in the curvature and yet, gives rise
to first order equations for the tetrad and the spin connection.   

\section{General Relativity as a gauge theory}

In this section we shall review some aspects of the vielbein --or
gauge-- formulation of general relativity.  The main point of this
section is to display the differences between gravity in odd and even
dimensions.   

\subsection{Poincar\'e translations vs. diffeomorphisms}
\label{Poincare/Diff}

General relativity in four and all even dimensions cannot be construed
as a `truly' local gauge theory\cite{Regge86}.  On the contrary, for
odd-dimensional spacetimes,  gravity can be written as a Chern-Simons
action for the Poincar\'e
group\cite{Achucarro-Townsend,Witten88,Chamseddine1}. [If the
cosmological constant is present, then the relevant group is the
(anti)-de Sitter group.]  The Poincar\'e group has generators $P_a$ and
$J_{ab}$ satisfying:

\begin{eqnarray}
~[P_a,P_b] &=& 0, \nonumber \\
~[P_a,J_{bc}] &=& \eta_{ab} P_c - \eta_{ac} P_b, \nonumber \\
~[J_{ab},J_{cd}] &=& \eta_{ac} J_{bd} - \eta_{bc} J_{ad} +\eta_{bd}
J_{ac} - \eta_{ad} J_{bc}. \label{poinc}
\end{eqnarray}
We define the connection for this group
\begin{equation} 
A= e^a P_a + \frac{1}{2} w^{ab} J_{ab}.
\end{equation} 

Let $\lambda$ be an arbitrary parameter with values in the Lie algebra
($\lambda = \lambda^a P_a +\frac{1}{2} \lambda^{ab}J_{ab}$). Under
the infinitesimal gauge
transformation $\delta A = \nabla \lambda$, $e^a$ and $w^{ab}$
transform as follows. 
\begin{eqnarray} 
Translations: && \mbox{\hspace{.5cm}} \delta e^a = D\lambda^a,
\mbox{\hspace{.5cm}} \delta w^{ab} = 0, \label{trans} \\
Rotations: && \mbox{\hspace{.5cm}} \delta e^a = \lambda^a_b e^b, 
\mbox{\hspace{.5cm}} \delta w^{ab} =- D \lambda^{ab}, 
\label{rotations}
\end{eqnarray}
where $D$ is the covariant derivative in the connection $w$. 

It might seem puzzling that even though the tetrad and the spin
connection carry a representation of the Poincar\'e group, the
Hilbert action constructed purely out of those fields, is not 
invariant under the Poincar\'e group in four dimensions. In fact, the
action 
\begin{equation}
I=\int \epsilon_{abcd} R^{ab}\mbox{\tiny $\wedge$} e^c 
   \mbox{\tiny $\wedge$} e^d
\end{equation}
is invariant under (\ref{rotations}) but not under (\ref{trans}).
The reason is simply that under (\ref{trans}) the action changes, modulo
boundary terms, by
\begin{equation}
\delta I = 2\int\epsilon_{abcd} R^{ab}\mbox{\tiny $\wedge$} T^c \lambda^d,
\end{equation}
which is not zero for arbitrary $\lambda$.

An alternative approach, often followed in  the supergravity literature,
is the so called 1.5 formalism. That is to set $T^a=0$ keeping only the
tetrad transformation in (\ref{trans}); the variation of $w^{ab}$ is
then calculated using the chain rule. This procedure brings the
diffeomorphism invariance into the scene because if $T^a=0$, the
variation (\ref{trans}) for the tetrad is equal -up to a rotation- to a
Lie derivative with a parameter $\xi^\mu=e^\mu_a \lambda^a$. 
In sum, the Hilbert action in 3+1 dimensions is invariant
under Lorentz rotations and diffeomorphisms, but not under
the local translations (\ref{trans}), generated by $P_a$.
 
A completely different situation is observed in 2+1 dimensions, in which
case the Hilbert action is linear in the dreibein field,
\begin{equation}
I_{2+1} = \int \epsilon_{abc} R^{ab}\mbox{\tiny $\wedge$} e^c. 
\label{2+1/action}
\end{equation}

It is straighforward to check, using the Bianchi identity, that the
variation of (\ref{2+1/action}) under (\ref{trans}) gives a boundary term
and hence, (\ref{trans}) is a symmetry of the 2+1 theory. As Witten has
pointed out, this simple fact has deep consequences \cite{Witten88}.
Indeed, in three dimensions, one can replace the diffeomorphism
invariance by a local Poincar\'e invariance whose constraint algebra is
a true Lie algebra. 

One may wonder if there exists an action invariant under (\ref{trans})
in higher dimensions. This generalization indeed exist and is given by   
\begin{equation} 
I_{2n+1} = \int
\epsilon_{a_1...a_{2n+1}} R^{a_1 a_2} 
 \mbox{\tiny $\wedge$} \cdots \mbox{\tiny $\wedge$} R^{a_{2n-1} a_{2n}}
\mbox{\tiny $\wedge$} e^{a_{2n+1}} 
\label{2n+1/action} 
\end{equation}
which, clearly, exists only in odd dimensions. The fact that
(\ref{2n+1/action}) can be written for odd dimensional manifolds only is
associated to the existence of Chern-Simons forms in those dimensions.
     
The key property of (\ref{2n+1/action}) is that it is linear in the
vielbein field rather than being linear in the curvature, as would be
the case for the Hilbert action. This fact  makes (\ref{2n+1/action})
invariant under the transformation (\ref{trans}), up to a boundary term.
Inspite of being non linear in the curvature, this action yields first
order differential equations for all the fields. This is not surprising
as (\ref{2n+1/action}) is a particular case of a Lovelock action
\cite{Lovelock}. $I_{2n+1}$ describes a Chern-Simons theory of
$ISO(D-1,1)$, obtained by contraction of $SO(D-1,2)$, and possesses
solutions with conical singularities \cite{BTZ2} analogous to those
found in 2+1 dimensions without cosmological constant
\cite{Deser-Jackiw-tHooft}. It should be mentioned here that the action
(\ref{2n+1/action}) has a propagating torsion\cite{BGH2} and therefore,
the 1.5 formalism is not applicable in this case. The main goal of this
paper is to describe the supersymmetric extension of the action 
(\ref{2n+1/action}).

\subsection{Super-Poincar\'e vs. supergravity}
\label{Poincare/sugra}

Supergravity if often referred to as the square root of General
Relativity \cite{CT-Tabensky} much in the same spirit as the Dirac
equation is the square root of the Klein-Gordon
equation. This is justified by the fact that the commutator of two
supersymmetry transformations gives a general change of coordinates
plus a rotation and another supersymmetry transformation. A concrete
example is the usual $N=1$ local supersymmetry transformations
\cite{PeterVN}
\begin{eqnarray} 
\delta e^a_\mu &=&\frac{1}{2} \bar \varepsilon \Gamma^a \psi_\mu,
 \nonumber \\ 
\delta \psi_\mu &=& D_\mu\varepsilon. \label{t2} 
\end{eqnarray} 
Here $\psi$ and $\varepsilon$ are Majorana spinors; $\bar\varepsilon$
is the Majorana conjugate $(\bar\varepsilon)_\alpha =
C_{\alpha_\beta}\varepsilon^\alpha$ satisfying $\bar\varepsilon
\Gamma^a \eta = - \bar \eta \Gamma^a \varepsilon$, and $D_\mu$ is the
covariant derivative in the spin connection.  It is then
straightforward to prove that the commutator of two transformations
with parameters $\varepsilon$ and $\eta$ acts on the tetrad as  
\begin{equation} 
[\delta_\varepsilon,\delta_\eta ]e^a_\mu =\frac{1}{2}
D_\mu(\bar\varepsilon \Gamma^a \eta). 
\label{a-e}
\end{equation}
The transformation (\ref{t2}) follows from the super-Poincar\'e algebra
when one considers the veilbein and the gravitino as the compensating
fields for local translations $(P_a)$ and supersymmetry transformations
($Q$), respectively.  It is not surprising therefore that the right hand
side of (\ref{a-e}) is a local translation --acting on the tetrad--,
with parameter $\lambda^a = \frac{1}{2} \bar\varepsilon \Gamma^a \eta$. 

The trouble with supergravity in $3+1$ dimensions is that the action    
{\em is not} invariant under local translations. Nevertheless,
transformation (\ref{a-e}) is still a symmetry of the action provided
the transformation of the spin connection preserves the torsion equation
$T^a=\frac{1}{2} \bar\psi \Gamma^a \mbox{\tiny $\wedge$} \psi$ 
(1.5 formalism). On the
surface defined by the torsion equation, the right hand side of
(\ref{a-e}) can be rewritten as  
\begin{equation}
D_\mu(\bar\varepsilon \Gamma^a
\eta) = {\cal L}_\xi e^a_\mu + \xi^\nu w^a_{b\nu} e^b_\mu - \xi^\nu
\psi_\nu \Gamma^a \psi_\mu.
\label{drs}
\end{equation}
The first term in the right hand side of (\ref{drs}) represents a
diffeomorphism with parameter $\xi^\mu=\bar\varepsilon
\Gamma^\mu \eta$ ($\Gamma^\mu e^a_\mu = \Gamma^a$), the second term is a
rotation with parameter $\xi^\mu w^a_{b\mu}$, and the third term is a
supersymmetry transformation with parameter $-\xi^\mu \psi_\mu$. 
Equation (\ref{drs}) shows that, as far as the tetrad is concerned, the
algebra of supersymmetry transformations closes.  But the fact that we
have used the torsion equation implies that the connection is no longer
an independent variable. On the contrary, its variation is given in
terms of $\delta e^a$ and $\delta \psi$, and {\it differs} from the form
dictated by group theory.  As a consequence, the local supersymmetry
algebra acting on the gravitino closes only on shell and 
auxiliary fields are required for its closure off shell.

A completely different situation is observed in the case of 2+1 gravity.
Although the above discussion applies to this case
\cite{Achucarro-Townsend}, there is an alternative route which leads more
directly to supergravity.  The key feature of 2+1 supergravity is that
--unlike 3+1 supergravity--, the action {\em is} invariant under local 
Poincar\'e transformations. This means that in the 2+1 theory it is not
necessary to express the right hand side of (\ref{a-e}) in terms of
diffeomorphisms; the commutator of two supersymmetry transformations
gives a local translation, which is a symmetry of the action as well.  

In other words, in 2+1 dimensions one can consider a {\em truly}
first order formalism in which the spin connection transforms
independently of the vielbein and gravitino 
just as dictated by group theory. In
particular, under supersymmetry, one can set
\begin{equation} 
\delta w^{ab}_\mu = 0
\end{equation}   
which, together with (\ref{t2}), is a symmetry of the 2+1 action. 

The simplicity of the 2+1 theory can be explicitly exhibited. The action
reads\cite{Achucarro-Townsend,Marcus-Schwarz/Deser-Kay}
\begin{equation}
I=\int ( \epsilon_{abc} R^{ab} \mbox{\tiny $\wedge$} e^c - 
\bar \psi\mbox{\tiny $\wedge$} D \psi),
\label{2+1/sugra}
\end{equation}
where $\psi$ is a two component Majorana spinor. [See the Appendix
for a summary of conventions.]

This action is invariant under Lorentz rotations,
\begin{eqnarray}
\delta\omega^{ab} &=& -D\lambda^{ab}, \mbox{\hspace{.5cm}} 
                 \delta e^a = \lambda^a_b e^a, \nonumber\\
\delta \psi &=& \frac{1}{4} \lambda^{ab}\Gamma_{ab} \psi; 
\label{2+1/rot}
\end{eqnarray}  
Poincar\'e translations, 
\begin{equation}
\delta w^{ab}=0,  \mbox{\hspace{.5cm}} 
 \delta e^a = D\lambda^a, \mbox{\hspace{.5cm}} \delta \psi = 0;
\label{2+1/transl}
\end{equation}  
and supersymmetry transformations,
\begin{equation}
\delta w^{ab}=0, \mbox{\hspace{.5cm}} 
\delta e^a = \frac{1}{2}\bar \varepsilon 
\Gamma^a \psi, \mbox{\hspace{.5cm}} 
\delta \psi = D \varepsilon.
\label{2+1/susy}
\end{equation}

The fields $e^a$, $w^{ab}$ and $\psi$ transform as components of a
connection for the super Poincar\'e group and therefore the
supersymmetry algebra implied by (\ref{2+1/rot})-(\ref{2+1/susy}) is
the super Poincar\'e Lie algebra. The invariance of the action
(\ref{2+1/sugra}) under the super Poincar\'e group should come as no
surprise because it is the Chern-Simons action for the
connection $A = e^aP_a + \frac{1}{2}w^{ab}J_{ab} +\bar Q \psi$, 
whose generators are $P_a,J_{ab}$ and $ Q^\alpha$. Indeed, the action 
(\ref{2+1/sugra}) can be written as\cite{Achucarro-Townsend} 
\begin{equation}
I = \int<A\mbox{\tiny $\wedge$} dA + \frac{2}{3} A 
\mbox{\tiny $\wedge$} A\mbox{\tiny $\wedge$} A>,
\label{2+1cs}
\end{equation}
where the bracket $< \cdots >$ stands for a properly normalized trace on
the algebra, with  $<J_{ab},P_c> = \epsilon_{abc}$ and
$<Q_\alpha,Q_\beta> = -iC_{\alpha\beta}$ are the only non-vanishing
traces.  [$C_{\alpha\beta}$ is the
charge conjugation matrix.]  

\section{Five dimensional Poincar\'e supergravity}
\label{5D/sugra}

We turn now to the supersymmetric version of the action
(\ref{2n+1/action}) in higher dimensions. To illustrate the ideas we
start with the five dimensional case which already contains the main
ingredients. The general case will be indicated
in the next section.  

\subsection{The Action}

The analog of the action (\ref{2+1/sugra}) in five dimensions has three
pieces;
\begin{equation}
I_{5/susy} = I_G + I_b + I_\psi
\label{5/superaction}
\end{equation}
where $I_G$ is the purely gravitational term, $I_b$ is a second
bosonic term needed by supersymmetry, and $I_\psi$ is the 
fermionic term. The explicit formulas are:
\begin{eqnarray}
I_G &=& \frac{1}{2}\int \epsilon_{abcdf} R^{ab} \mbox{\tiny $\wedge$} 
          R^{cd} \mbox{\tiny $\wedge$} e^f, \nonumber\\
I_b &=&  \int R^{ab}\mbox{\tiny $\wedge$} R_{ab} \mbox{\tiny $\wedge$}
b, \nonumber\\
I_\psi &=&  -\int R^{ab} \mbox{\tiny $\wedge$} (\bar \psi \Gamma_{ab}
 \mbox{\tiny $\wedge$} D\psi +
            D \bar \psi \Gamma_{ab} \mbox{\tiny $\wedge$} \psi). 
\end{eqnarray}
Here the gravitino field is a Dirac spinor 1-form, and $b$ is a 1-form
Lorentz pseudo-scalar. The form of $I_b$ is dictated by the
transformation of $I_G$ and $I_\psi$ under supersymmetry. [See section
\ref{Integrability} for an alternative way to see this through the
integrability conditions of the classical equations of motion.]

Since we are interested in the geometrical aspects of this theory only,
we have set all the coupling constants equal to one.  The Lorentz
covariant derivative $D$ in the spinorial representation is given by
$D\psi=d\psi + \frac{1}{4}w^{ab}\Gamma_{ab} \mbox{\tiny $\wedge$}
\psi$.  We work with Dirac
spinors in order to avoid the dimensional dependence of Majorana
spinors which only exist -on a Minkwoskian signature- in
dimensions 2,3,4 $mod$ 8. All the main results of this paper carry
through for Majorana spinors when they exist. The Dirac
conjugate is defined as $\bar \psi = \psi^{\dag} \Gamma_0 $ and 
in the action, $\psi$ and $\bar \psi$ are varied independently. [See 
the Appendix.]

Besides local Lorentz rotations, the above action is invariant under
the Abelian translations,    
\begin{eqnarray}
   \delta e^a &=& D\lambda^a, \nonumber\\
\delta w^{ab} &=& 0, \nonumber\\
     \delta b &=& d\rho, \nonumber\\
  \delta \psi &=& 0,   
\label{super/transl}
\end{eqnarray}
where $\lambda^a$ is a 0-form Lorentz vector and $\rho$ is a 0-form
Lorentz pseudo-scalar. The invariance of (\ref{5/superaction}) under
(\ref{super/transl}) follows directly from
the Bianchi identity. The action (\ref{5/superaction}) is also
invariant under supersymmetry transformations,
\begin{eqnarray}
   \delta e^a &=& -i (\bar \varepsilon \Gamma^a \psi - \bar \psi \Gamma^a
                   \varepsilon),  \nonumber\\
\delta w^{ab} &=& 0, \nonumber \\
 \delta b &=& \bar \varepsilon \psi - \bar \psi \varepsilon, \nonumber\\
\delta \psi &=& D\varepsilon. 
\label{super/transf}
\end{eqnarray}

The proof of the invariance of (\ref{5/superaction}) under
(\ref{super/transf}) is straightforward. An important test  
of the consistency of (\ref{super/transf})  is the
fact that the commutator of two supersymmetry transformations
gives local translation
(\ref{super/transl}).  Indeed, if $\delta_\varepsilon$ and $\delta_\eta$
are supersymmetry transformations with parameters $\varepsilon$ and
$\eta$, we have,
\begin{eqnarray}
~[\delta_\varepsilon,\delta_\eta] e^a &=& -iD(\bar \varepsilon 
\Gamma^a \eta -
\bar \eta \Gamma^a \varepsilon), \nonumber\\
~[\delta_\varepsilon,\delta_\eta] w^{ab} &=& 0, \nonumber\\
~[\delta_\varepsilon,\delta_\eta] b &=& d(\bar \varepsilon \eta - 
\bar \eta \varepsilon), \nonumber \\ 
~[\delta_\varepsilon,\delta_\eta] \psi &=& 0 .
\label{super/algebra0}
\end{eqnarray}

The symmetries of the action (\ref{5/superaction}) are generated by the
local super-Poincar\'e generators $P_a$, $J_{ab}$, $Q^{\alpha}$,
$\bar{Q}_{\alpha}$, plus the Abelian generator $K$, responsible for the
non-zero transformation of $b$ in (\ref{super/transl}). These
generators form an extension of the super-Poincar\'e algebra whose only
non-vanishing (anti-)commmutator is
\begin{eqnarray}
\{ Q^\alpha,\bar Q_\beta\} &=& -i(\Gamma^a)^\alpha_\beta P_a +
    \delta^\alpha_\beta K,
\label{QQ}
\end{eqnarray}
plus the Poincar\'e algebra.  The commutators of $K$, $Q^\alpha$, and
$\bar Q_\alpha$ with the Lorentz generators can be read off from their
tensor character. 

Note that $K$ commutes with all the generators in the algebra and
therefore, it is as a central charge in the super Poincar\'e algebra. 
This is, however, a peculiarity of five dimensions. For other odd
dimensions, the generator $K$ is a completely antisymmetric tensor of
fifth rank, which has a non-vanishing commutator with the Lorentz
generator.
 
\subsection{Chern-Simons formulation}

The fact that the symmetries of the action (\ref{5/superaction}) close
without the need of any auxiliary fields strongly suggest that
(\ref{5/superaction}) may be written as a Chern-Simons action for the
supergroup.  In this section we prove that this is indeed the case. 

Consider a connection $A$ for the superalgebra found in the last
section,
\begin{equation}
A = e^a P_a + \frac{1}{2} w^{ab}J_{ab} + b K +\bar \psi_\alpha
Q^\alpha - \bar Q_\alpha \psi^\alpha.
\label{A}
\end{equation}
The super-curvature $F=dA+A\mbox{\tiny $\wedge$} A$ is then found 
by direct application of (\ref{poinc}) and (\ref{QQ})
\begin{eqnarray}
F &=& F^A G_A \nonumber \\
&=& \tilde T^a P_a + \frac{1}{2} R^{ab}J_{ab} + \tilde F K + 
D \bar \psi_\alpha
Q^\alpha - \bar Q_\alpha D \psi^\alpha.  
\label{F}
\end{eqnarray}
Here $\tilde T^a := T^a -i \bar \psi \Gamma^a 
\mbox{\tiny $\wedge$} \psi$ and 
$\tilde F := db + \bar \psi \mbox{\tiny $\wedge$} \psi$; $T^a$ is the 
torsion 2-form, and $R^{ab}$ is the 2-form Lorentz curvature.  

We recall that a Chern-Simons
Lagrangian in five dimensions, $L_g$, is defined by the relation
\begin{equation}
g_{ABC} F^A \mbox{\tiny $\wedge$} F^B \mbox{\tiny $\wedge$} F^C = dL_g, 
\end{equation}
where the trilinear form $g_{ABC} \equiv <G_A,G_B,G_C>$ 
is an invariant tensor of the Lie algebra with generators $G_A$. 
Different choices of the invariant tensor $g_{ABC}$ give different
five dimensional Lagrangians $L_g$. 

To prove that (\ref{5/superaction}) is a Chern-Simons action we need to
find an invariant tensor such that $L_g$ is equal, up to a total
derivative, to the Lagrangian in (\ref{5/superaction}). This tensor
indeed exists and is given by,
\begin{eqnarray}
<J_{ab},J_{cd},P_e> &=& \epsilon_{abcde},  \nonumber \\
<J_{ab},J_{cd},K> &=& \eta_{ac} \eta_{bd}- \eta_{ad} \eta_{bc} 
                                              \nonumber   \\
<Q^\alpha, J_{ab}, \bar Q_\beta > &=&-2(\Gamma_{ab})^\alpha_\beta. 
\label{inv/ten3}
\end{eqnarray}
It is straightforward to prove that, up to a total derivative and an
overall factor, the
5-form $L_g$ associated to the above tensor is equal to the
supersymmetric Lagrangian in (\ref{5/superaction}). 

>From the above result it is now evident that the action is invariant
under supersymmetry transformations up to a total derivative.  All the
symmetry transformations of the action can now be collected together in
the form   $\delta A = \nabla \lambda$  where $\nabla$ is the covariant
derivative of the supergroup and $\lambda$ is a zero-form Lie
algebra-valued vector in the adjoint representation. From this formula
it is also evident that the algebra of supersymmetry transformations 
closes as dictated by group theory, 
\begin{equation} 
~[\delta_{\lambda}, \delta_{\eta} ] A =
\delta_{[\lambda,\eta]} A. 
\end{equation}  
The Chern-Simons action
(\ref{5/superaction}) can be obtained from the action found in
\cite{Chamseddine2} by an appropriate Wigner-Inonu contraction.   The
closure of the supersymmetry algebra, however, was not mentioned in
\cite{Chamseddine2}. In the next section we prove that the above scheme
is not exclusive of the three and five dimensional theories but 
it can be extended to any odd-dimensional spacetime.    

\section{The general case}

\label{D/sugra}

In this section we show how the results of the previous sections 
are generalized to any odd-dimensional manifold.  In order to simplify
the notation we introduce the symbol $R_{abc}$ defined by,
\begin{equation}
R_{abc} := \epsilon_{abc a_1 \cdots a_{D-3}}
R^{a_1 a_2} \mbox{\tiny $\wedge$} \cdots \mbox{\tiny $\wedge$} 
R^{a_{D-2} a_{D-3}}.
\label{R/symbol}
\end{equation}

Just as in the five dimensional case, the supersymmetric
action in dimension $2n+1$ has three terms;
\begin{equation}
I_{2n+1/susy} = I_G + I_b + I_{\psi}
\label{2n+1/susy}
\end{equation}
where the bosonic `geometric' term $I_G$ is given by,
\begin{equation}
I_G = \int R_{abc}\mbox{\tiny $\wedge$} R^{ab} \mbox{\tiny $\wedge$}e^c.
\label{2n+1/action'}
\end{equation}   
$I_{b}$ is a second bosonic term involving a fifth-rank 1-form field,
$b^{abcde}$,
\begin{equation} 
I_{b}= -\frac{1}{6} \int R_{abc} \mbox{\tiny $\wedge$} R_{de} 
\mbox{\tiny $\wedge$} b^{abcde}. 
\end{equation}
(Note that this term vanishes in three dimensions and in five dimensions
$b^{abcde}$ is a Lorentz pseudo-scalar.) Finally, 
the fermionic part is
\begin{equation}
I_{\psi}=\frac{i}{3}\int R_{abc} \mbox{\tiny $\wedge$} 
(\bar \psi\Gamma^{abc}\mbox{\tiny $\wedge$} D\psi 
+ D\bar\psi\Gamma^{abc} \mbox{\tiny $\wedge$} \psi).
\end{equation}
Each term in the action (\ref{2n+1/susy}) is independently invariant
under local Lorentz transformations. The complete action is invariant 
under the Abelian translations, 
\begin{eqnarray}
\delta e^a &=& D\lambda^a, \mbox{\hspace{.5cm}} \delta w^{ab}=0, 
\mbox{\hspace{.5cm}} \delta b^{abcde} =
D\rho^{abcde},  \nonumber\\
\delta \psi &=& 0,  
\label{abelian/2n+1}
\end{eqnarray}
and supersymmetry transformations,
\begin{eqnarray}
\delta e^a = -i(\bar \varepsilon \Gamma^a \psi-\bar\psi\Gamma^a
\varepsilon), \mbox{\hspace{.5cm}} \delta w^{ab} =0, \nonumber \\ 
\delta b^{abcde} = -i(\bar \varepsilon \Gamma^{abcde}\psi -h.c.), 
\mbox{\hspace{.5cm}} \delta \psi = D\varepsilon,   
\label{susy.2n+1} 
\end{eqnarray}
where $D$ represents the Lorentz covariant derivative.

The proof of the invariance of (\ref{2n+1/susy}) under supersymmetry
 transformations is straighforward. 
One starts by varying the fermionic part. Up to a
boundary term one easily obtains
\begin{equation}
\delta I_\psi=\frac{i}{12}\int R_{abc}\mbox{\tiny $\wedge$} 
R_{de}\mbox{\tiny $\wedge$} (\bar \psi 
\{\Gamma^{abc},\Gamma^{de}\} \epsilon - c.c. ).
\label{proof/1}
\end{equation}
Using the formula (\ref{a2}) of the Appendix, we find that
(\ref{proof/1}) has a term proportional to a 5-rank Dirac matrix
$\Gamma^{abcde}$ plus a term proportional to a Dirac matrix $\Gamma^a$.
It is direct to see that the first term is cancelled by the variation
of $I_b$ while the second term is cancelled by the variation of $I_G$. 

As in the lower dimensional cases ($D=3,5$), the commutator of two
supersymmetry transformations gives a local Abelian translation
(\ref{abelian/2n+1}). Thus, the supersymmetric extension of the
Poincar\'e algebra that leaves the action invariant has generators $G_A
=[P_a, J_{ab}, K_{abcde}, Q^{\alpha}, \bar Q_{\alpha}]$. The only
non-vanishing (anti-)commutator is 
\begin{equation}
\{Q^\alpha,\bar Q_\beta \} = -i(\Gamma^a)^\alpha_\beta P_a - 
      i (\Gamma^{abcde})^\alpha_\beta K_{abcde},
 \label{U}
\end{equation}
plus the Poincar\'e algebra. The commutators of the 
$Q, \bar Q$ and
$K$ with the Poincar\'e generators can be read from their
tensorial character.  

The action $I_{2n+1/susy}$ is also a Chern-Simons action. The connection
now is
\begin{equation}
A = e^a P_a + \frac{1}{2} w^{ab}J_{ab} + b^{abcde} K_{abcde} +\bar
\psi_\alpha Q^\alpha - \bar Q_\alpha \psi^\alpha,
\end{equation}
and the Lagrangian is defined by $<F\mbox{\tiny $\wedge$} \cdots 
\mbox{\tiny $\wedge$} F> = d L_{2n+1}$
where the invariant $(n+1)$ multilinear form $<\cdots>$ is defined by 
\begin{eqnarray}
<J_{a_1a_2} \cdots J_{a_{D-2}a_{D-1}} P_{a_{D}}> &=& \epsilon_{a_1 \cdots
a_D}, \nonumber\\
<J_{a_1a_2} \cdots J_{fg} K_{abcde}> &=&
-\frac{1}{12}\epsilon_{a_1 \cdots a_{D-3} a b c} \eta_{[fg][de]},
\nonumber \\
<Q J_{a_1a_2} \cdots J_{a_{D-4}a_{D-3}} \bar Q> &=& 2i^n \Gamma_{a_1
\cdots a_{D-3}} 
\end{eqnarray} 
(the remaining brackets are zero). This completes the construction of
the $(2n+1)$-dimensional Chern-Simons action for supergravity.

\section{Integrability of the equations of motion.}
\label{Integrability}

A remarkable feature of supersymmetric theories in general and 
supergravity in particular is the fact that the integrability
conditions for the fermionic field equations are the bosonic equations.
The study of the integrability conditions of the fermionic equation in
our model sheds some light on the role of the bosonic field $b^{abcde}$.

In the notation introduced in Sec. IV, the fermionic field equations are
\begin{equation}
R_{abc}\Gamma^{abc}\mbox{\tiny $\wedge$} D \psi =0,
\label{fer/eq}
\end{equation}
and similarly for the Dirac conjugate spinor. Taking the covariant
derivative of (\ref{fer/eq}) we find the integrability condition,
\begin{equation}
R_{abc}\mbox{\tiny $\wedge$} R^{de} \Gamma^{abc} 
\Gamma_{de}\mbox{\tiny $\wedge$} \psi =0.
\label{fer/integ}
\end{equation}
This equation should be satisfied for any $\psi$.  Using 
elementary properties of the Dirac matrices we obtain the following
equations for the bosonic fields,
\begin{eqnarray}
 R_{abc} \mbox{\tiny $\wedge$} R^{ab} \Gamma^c = 0,\label{cons1}  \\ 
 R_{abc} \mbox{\tiny $\wedge$} R_{de} \Gamma^{abcde} =0.\label{cons2}
\end{eqnarray}
Eq. (\ref{cons1}) is the equation of motion for the vielbein field,
while Eq. (\ref{cons2}) is the equation of motion for the $b-field$. 
Had we not included $b$, supersymmetry would not have been achieved and
the integrability conditions would not have been satisfied. [This does
not rule out a Lagrangian without the $b$ field. However, the fermionic
equations for such a theory would impose additional equations on the
bosonic fields.] 


\section{Comments and prospects}

We have shown in this note that the successful methods used in three
dimensions to construct supersymetric extensions of general relativity
can be generalized to any odd-dimensional spacetime. We have restricted
ourselves, however, to Poincar\'e supergravity. The full anti-de Sitter
extension remains an open problem. [In five dimensions, a Chern-Simons
action for anti-de Sitter supergravity has been known for some time
\cite{Chamseddine2}. That action reduces to the action considered here
after a proper contraction is performed.] 

There are good reasons to seek a full anti-de Sitter Chern-Simons
formulation of supergravity. First, the bosonic Lagrangian in the
Poincar\'e case does not contain the Hilbert term thus making the
contact with four dimensional theories rather obscure
\cite{Chamseddine1}.  Secondly, the Poincar\'e theory in odd dimensions
does not possess black hole solutions while the anti-de Sitter theory
does \cite{BTZ2}.  

In principle, a Chern-Simons anti-de Sitter supergravity can be
constructed from the knowledge of the associated supergroup and an
invariant tensor only (finding the invariant tensor, however, may prove
to be a non-trivial task).  In five dimensions, the relevant supergroup
is $SU(2,2|1)$ \cite{Nahm} while in the important example of eleven
dimensions the supergroup is $OSp(32|1)$ \cite{Holten-Proeyen}.  As the
spacetime dimension increases, one faces a growing multiplicity of
choices for the invariant tensor. To illustrate this issue, consider the
problem of classifying all the invariants that can be constructed  out
the Lorentz curvature in a given dimension\cite{Jano}. In four
dimensions we only have, 
\begin{equation}
\epsilon_{abcd}R^{ab}\mbox{\tiny $\wedge$} R^{cd}, \mbox{\hspace{.5cm}} 
R^{ab} \mbox{\tiny $\wedge$} R_{ab}, 
\end{equation} 
while in eight dimensions we have, 
\begin{eqnarray} 
\epsilon_{abcdefgh} R^{ab}\mbox{\tiny $\wedge$} R^{cd} 
\mbox{\tiny $\wedge$} R^{ef}\mbox{\tiny $\wedge$}
R^{gh},  && \mbox{\hspace{.5cm}} R^{ab} 
\mbox{\tiny $\wedge$} R_{ab} \mbox{\tiny $\wedge$} 
R^{cd} \mbox{\tiny $\wedge$} R_{cd}, \nonumber\\
R^{ab} \mbox{\tiny $\wedge$} R_{bc} \mbox{\tiny $\wedge$} 
R^{cd} \mbox{\tiny $\wedge$} R_{da}&& . 
\end{eqnarray}  
Of course, all the above scalars define Chern-Simons Lagrangians in 
dimension five and seven respectively. A similar proliferation of
scalars appears in  supergravity.  A good candidate for the right theory
could be a linear combination of all possible invariants such that,
under an appropriated Wigner-Inonu contraction, reduces to the
Poincar\'e theories studied here.

The particular case of eleven dimensions seems to be particulary suited
to admit an anti-de Sitter Chern-Simons formulation. As shown in
\cite{Holten-Proeyen}, the super anti-de Sitter group is $OSp(32|1)$. A
natural basis for the Lie algebra of $Sp(32)$ is given by the Dirac
matrices $\Gamma_a, \Gamma_{ab}, \Gamma_{abcde}$, and this basis is
easily extended to expand the superalgebra of $OSp(32|1)$. Thus, the
supergroup $OSp(32|1)$ naturally accommodates the field content of the
Poincar\'e Chern-Simons supergravity considered here. One could expect,
therefore, that a Chern-Simons Lagrangian for the supergroup $OSp(32|1)$
in eleven dimensions should reduce to the supersymmetric action
(\ref{2n+1/susy}) upon contraction.  For example, 
is easy to check that the superalgebra obtained in
Sec. (\ref{D/sugra}) can 
be obtained from $OSp(32|1)$ by a Wigner-Inonu contraction:
\begin{eqnarray}
G_a &&\rightarrow \lambda^{-1} P_a, \nonumber \\
G_{ab} &&\rightarrow J_{ab},  \nonumber \\
G_{abcde} &&\rightarrow \lambda^{-1} K_{abcde}, \nonumber \\
Q^{\alpha} &&\rightarrow \lambda^{-\frac{1}{2}} Q^{\alpha}.
\label{W-I}
\end{eqnarray}
At the level of the Lagrangian, however, the problem is more complicated.
Due to the ambiguity in the choice of the invariant tensors and the
large number of terms in the super anti-de Sitter Chern-Simons action,
it is a non-trivial problem to find an expression such that, under
contraction, reduces to the action considered in this paper\cite{BTrZ2}.

Finally we mention that the -Poincar\'e- supersymmetric Chern-Simons 
actions found in this paper are not the only possibilities. 
In  dimensions
greater than five, the fermionic Lagrangian accepts other
Poincar\'e-invariant terms that give rise to other supergravities.
 For example, in eleven
dimensions one can add to the fermionic lagrangian the term,
\begin{equation}
 [R^{ab}  \mbox{\tiny $\wedge$} R_{ab}]^2  \mbox{\tiny $\wedge$}
 \bar \psi \mbox{\tiny $\wedge$} D\psi  
\label{last}
\end{equation}
This term, however, requires extra bosonic fields to respect 
supersymmetry. This is easily seeing by 
studying the integrability conditions generated by 
(\ref{last}). One finds the equation over the bosonic fields,
\begin{equation}
 [R^{ab}  \mbox{\tiny $\wedge$} R_{ab}]^2  \mbox{\tiny $\wedge$}
R^{cd} =0. \label{last2} 
\end{equation}
Thus, consistency requires an extra bosonic term  in the Lagrangian 
of the form 
$ [R^{cd}  \mbox{\tiny $\wedge$} R_{cd}]^2  \mbox{\tiny $\wedge$}
R^{ab}  \mbox{\tiny $\wedge$} c_{ab}$ which involves
the 1-form $c_{ab}$. Varying this term with respect to $c_{ab}$ gives
(\ref{last2}).  Thus, the integrability condition is satisfied and the
action is supersymmetric. 
A complete classification of all possible fermionic Lagrangians for a
given dimension and their corresponding supersymmetry algebras
is beyond the scope of this work.  We would like to
point out, however, that the method outlined here seems to provide a
simple way to generate extensions of the super-Poincar\'e algebra
involving extra bosonic fields.


\appendix

\section{Gamma matrices}

The Clifford algebra in $D=2n+1$ dimensions with Minkowskian signature
can be generated by a set of $2^n\times 2^n$ matrices; the unit $I$ and
$D$ matrices $\Gamma^a$, satisfying
$\{\Gamma^a,\Gamma^b\}$=$2\eta^{ab}I$, where $a, b,..., =1,2,...,2n+1$
and $\eta^{ab}$=$diag(-,+,\cdots,+)$. In this signature,
$\Gamma^{\dag}_a = \Gamma_0 \Gamma_a \Gamma_0$. 

It is always possible to find a
representation of $\Gamma^a$ matrices in which 
\begin{equation}
\Gamma := \Gamma^1\Gamma^2\cdots\Gamma^D = (-i)^{n+1}I.
\end{equation} 
We define $\Gamma^{a_1\cdots a_p}$ as the totally antisymmetric
product of gamma matrices,
\begin{equation}
\Gamma^{a_1 \cdots a_p} = \frac{1}{p!} \sum_{\sigma} sgn(\sigma)
\Gamma^a_{\sigma(1)} \cdots \Gamma^a_{\sigma(p)}, 
\end{equation}
Two useful formulas implicitly used in the text are:
\begin{equation}
\Gamma_{a_1\cdots a_{D-3}} =
-\frac{(-i)^{n+1}}{3!}\epsilon_{a_1\cdots a_{D-3}abc}\Gamma^{abc}
\label{a1}
\end{equation}
\begin{equation}
\frac{1}{2}\{\Gamma^{abc},\Gamma^{de}\}= \Gamma^{abcde}
-[\eta^{[ab][de]}\Gamma^c + perm(abc)],
\label{a2}
\end{equation}
where $\eta^{[ab][de]} = \eta^{ad}\eta^{be}-\eta^{ae}\eta^{bd}$.  \\
\\

\noindent {\large {\bf Acknowledgments}}\\

The authors are grateful to M. Contreras, C. Mart\'{\i}nez and,
especially, C. Teitelboim for helpful discussions. This work was
supported in part by grants 1930910 and 2950047 from FONDECYT (Chile),
27-953/ZI-DICYT  (University of Santiago) and grant No. PG-023-95 of
Departamento de Postgrado y Post\'{\i}tulo (University of Chile). M.B.
was partially supported by a grant from Fundaci\'on Andes. R.T. thanks
CONICYT for financial support. Institutional support to CECS from a
group of Chilean private companies (COPEC, CGEI, Empresas CMPC, ENERSIS,
MINERA LA ESCONDIDA, IBM and XEROX-Chile) is also acknowledged.


\begin{references}

\bibitem{Green-Schwarz-Witten} For a complete treatment see,
             M.B. Green, J.H. Schwarz and E. Witten, 
    \underline{Superstrings}, {\em Cambridge University Press}, 1987. 

\bibitem{Cremmer-Julia-Scherk} E. Cremmer, B. Julia and J. Scherk, 
{\em Phys. Lett.} {\bf B76}, 409 (1978).

\bibitem{Witten88} E. Witten, {\em Nucl. Phys.} {\bf B 311}, 46 (1988).

\bibitem{Achucarro-Townsend} A. Ach\'ucarro and P.K. Townsend, 
{\em Phys. Lett.} {\bf B180}, 89 (1986). 

\bibitem{Chamseddine2} A.H. Chamseddine, 
{\em Nucl. Phys.} {\bf B 346}, 213 (1990).

\bibitem{Chamseddine1} A.H. Chamseddine, 
{\em Phys. Lett.} {\bf B233}, 291 (1989).

\bibitem{BTZ2} M. Ba\~nados, C. Teitelboim and J.Zanelli,
{\em Phys. Rev.} {\bf D49}, 795 (1994).  See also, \underline{JJ Giambiagi
Festschriff}, H. Falomir, R. Gamboa, P. Leal and F. Shaposnik eds. ({\em 
World Scientific}, Singapore, 1991).  

\bibitem{BTZ} M. Ba\~nados, C. Teitelboim and J.Zanelli,
{\em Phys. Rev. Lett} {\bf 69}, 1849 (1992).

\bibitem{BGH} M. Ba\~nados, L.J. Garay and M. Henneaux, {\em Phys.
Rev.D},  in press.

\bibitem{Regge86} T. Regge, {Phys. Rep.} {\bf 137}, 31 (1986).

\bibitem{Lovelock} D. Lovelock, {\em J. Math. Phys.} {\bf 12}, 498 (1971).

\bibitem{Deser-Jackiw-tHooft} S. Deser, R. Jackiw and G. 't Hooft,
{\em Ann. Phys.} {\bf 152}, 220 (1984).

\bibitem{BGH2} M. Ba\~nados, L.J. Garay and M. Henneaux, in preparation. 

\bibitem{CT-Tabensky} C. Teitelboim, {\em Phys. Rev. Lett.} {\bf 38}, 1106
(1977); C. Teitelboim and R. Tabensky, 
               {\em Phys. Lett.} {\bf 69 B}, 240 (1977).

\bibitem{PeterVN} For a review see P. van Nieuwenhuizen, 
{\em Phys. Rep.} {\bf 68}, 4 (1981).

\bibitem{Marcus-Schwarz/Deser-Kay} For earlier work on three dimensional
supergravity see N. Marcus and J.H. Schwarz,
{\em Nucl. Phys.} {\bf B228}, 145 (1983); S. Deser and J.H. Kay, {\em
Phys. Lett.} {\bf B120}, 97 (1983).  

\bibitem{Nahm} W. Nahm, {\em Nucl. Phys.} {\bf B135}, 149 (1978).

\bibitem{Holten-Proeyen} J. W. van Holten and A. Van Proeyen, {\em J.
Phys. }{\bf A15}, 3763 (1982). 

\bibitem{Jano} A. Mardones and J. Zanelli, {\em Class. \& Quan. Grav.}
               {\bf 8}, 1545 (1991). 

\bibitem{BTrZ2} M. Ba\~nados, R. Troncoso and J. Zanelli, in preparation. 


\end{references}
\end{document}